\documentclass[twocolumn]{aastex63}
\usepackage[utf8]{inputenc}
\usepackage{bm}
\usepackage{textcomp}
\usepackage{graphicx}
\usepackage{amssymb}
\usepackage{amsmath}
\usepackage{gensymb}
\usepackage{booktabs}

\received{}
\revised{}
\accepted{}

\shorttitle{P-CORONA}
\shortauthors{Supriya et al.}

\begin{document}
\title{P-CORONA: A New Tool for Calculating the Intensity and Polarization of Coronal Lines in \\3D Models of the Solar Corona}

\correspondingauthor{Supriya Hebbur Dayananda}
\email{supriya@iac.es}

\author[0000-0003-2752-7681]{Supriya Hebbur Dayananda}
\affiliation{Instituto de Astrofísica de Canarias, E-38205 La Laguna, Tenerife, Spain}
\affiliation{Departamento de Astrofísica, Facultad de Física, E-38206 La Laguna, Tenerife, Spain}

\author[0000-0002-4930-7754]{Ángel de Vicente}
\affiliation{Instituto de Astrofísica de Canarias, E-38205, La Laguna, Tenerife, Spain}
\affiliation{Departamento de Astrofísica, Facultad de Física, E-38206, La Laguna, Tenerife, Spain}

\author[0000-0003-1465-5692]{Tanausú del Pino Alemán}
\affiliation{Instituto de Astrofísica de Canarias, E-38205, La Laguna, Tenerife, Spain}
\affiliation{Departamento de Astrofísica, Facultad de Física, E-38206, La Laguna, Tenerife, Spain}

\author[0000-0001-5131-4139]{Javier Trujillo Bueno}
\affiliation{Instituto de Astrofísica de Canarias, E-38205, La Laguna, Tenerife, Spain}
\affiliation{Departamento de Astrofísica, Facultad de Física, E-38206, La Laguna, Tenerife, Spain}
\affiliation{Consejo Superior de Investigaciones Científicas, Spain}

\author[0000-0003-3958-9935]{Nataliia G. Shchukina}
\affiliation{Instituto de Astrofísica de Canarias, E-38205, La Laguna, Tenerife, Spain}
\affiliation{Main Astronomical Observatory, National Academy of Sciences, 03143 Kyiv, Ukraine}

\begin{abstract}  
 The critical need to study the magnetic field in the solar corona is highlighted by recent observational facilities, such as DKIST and Aditya-L1. A powerful tool for probing the magnetism of the solar corona is forward modeling of the intensity and polarization of coronal emission lines in three-dimensional (3D) magnetohydrodynamic models. Here we present \mbox{P-CORONA}, a new spectral synthesis code designed to calculate the intensity and polarization of coronal lines in 3D models of the solar corona, taking into account the symmetry breaking induced by magnetic and velocity fields. \mbox{P-CORONA} allows the calculation of the on-disk and off-limb intensity and polarization of forbidden and permitted coronal lines, thus facilitating a wide range of investigations. Applying the quantum theory of atom-photon interactions, \mbox{P-CORONA} accounts for the spectral line polarization caused by anisotropic radiation pumping and the Hanle and Zeeman effects, making it a valuable
 tool for investigating coronal magnetic fields. This paper details the code’s theoretical formulation, the implementation, and illustrative results of calculations in different 3D coronal models (MURaM and Predictive Science Inc.), including the
 impact of the Zeeman effect from the transverse magnetic field component on selected coronal forbidden lines. 
 \mbox{P-CORONA} is now accessible to the research community on GitLab and Zenodo, providing a resource to facilitate research aimed at advancing our understanding of coronal magnetism and dynamics.
\end{abstract}


\section{Introduction} \label{sec:intro}
The solar corona plays a key role in shaping the heliosphere and influencing space weather. Understanding its complex dynamics is essential for predicting solar events that can impact space and Earth-based technologies. Of fundamental importance to understand the physics of this very hot rarefied plasma is to extract information about its magnetic field, which governs the coronal dynamics. The information on coronal magnetic fields is encoded in the polarization of spectral lines formed in the corona, both forbidden and permitted lines. Despite significant advances in observational techniques, interpreting the intensity and polarization of coronal spectral lines remains challenging. Recent observational efforts, such as those by the Upgraded Coronal Multi-channel Polarimeter \citep[UCoMP;][]{2016JGRA..121.8237L,2021AGUFMSH15G2089T} and the Daniel K. Inouye Solar Telescope (DKIST) using the Cryogenic Near-Infrared Spectropolarimeter \citep[CryoNIRSP;][]{2014SPIE.9147E..07E,2020SoPh..295..172R,2023SoPh..298....5F}, have provided measurements of the intensity and polarization of various infrared (IR) lines in the corona. However, extracting physical parameters, particularly the magnetic field, from these observations is complex. This complexity arises from various physical processes that influence the generation of polarization in spectral lines, as well as from the fact that the observed signal results from the integrated emission along the line-of-sight (LOS).

The forward modeling technique involves computing the intensity and polarization of coronal spectral lines under specific physical conditions to generate synthetic profiles, which can then be compared with spectropolarimetric observations \citep[e.g., the review by][]{2022ARA&A..60..415T}. The main physical processes influencing coronal spectral lines include photoexcitation, collisional excitation, Doppler dimming and brightening, and the impact of the magnetic field through the Hanle and Zeeman effects. Additionally, the active regions on the photosphere \citep[e.g.,][]{2021SoPh..296..166S} and Thomson scattering \citep[e.g.,][]{2017ApJ...838...69L} can also have an impact on coronal spectral profiles. 

There have been various research efforts to develop forward modeling codes for computing the intensity and polarization of coronal spectral lines.
Efforts in this direction include the Coronal Line Emission \citep[CLE;][]{2001ASPC..236..503J, 2006ApJ...651.1229J}, code to synthesize the Stokes profiles of coronal forbidden lines. The Python package for Coronal Emission Line Polarization (pyCELP), developed by \cite{2020SoPh..295...98S}, is a Python-based version of the CLE code. In developing this version, the authors fixed a bug concerning the computation of collisional rates and included non-dipole radiative transitions. However, these codes were developed for the specific case of spectral lines in the so-called saturated Hanle regime. In this regime, the spectral lines are only sensitive to the magnetic field orientation but not to its magnitude. Forbidden lines, usually magnetic dipole transitions with relatively small transition probabilities, are in this regime. However, some permitted lines formed in the corona, which result from electric dipole transitions, can be sensitive to both the strength and orientation o
f the magnetic field, depending on their specific transition probabilities and the local magnetic field conditions. 
An exception to this are the Fe {\sc x} lines at 174.5 \AA\ and 177 \AA, whose predicted  
linear polarization signals (due to an interesting physical mechanism) 
are sensitive only to the magnetic field 
orientation \citep[see][]{2009ASPC..405..423M}. We refer the readers to the review papers by \cite{Casini+2017,TrujilloBueno+2017, 2022ARA&A..60..415T} for further details on these topics. 
 FORWARD by \citet{2016FrASS...3....8G} is another forward modeling tool developed 
 to calculate the Stokes parameters of permitted and forbidden lines, along with the capability of modeling the white-light corona and the radio polarization signals. It is a collection of a set of different codes, including the above-mentioned CLE code for forbidden lines. Different physical mechanisms influencing the spectral line formation are included or excluded depending on the spectral line of interest. Recently, \citet{2024ApJ...977...97M} presented spectropolarimetric calculations using the \mbox{ScatPolSlab} code to emphasize that the Hanle effect in the He {\sc i} 10830 \AA{} line can serve as a robust diagnostic of the magnetic field in erupting prominences. \citet{2022Ap&SS.367..125F} used the HanleCLE code to simulate Hanle effect signatures in UV resonance lines, exploring their diagnostic potential for detecting and mapping magnetic fields in stellar winds and circumstellar environments. Previously, there were efforts by \citet{2002Natur.415..403T} on the mode
ling of the Hanle effect in the He {\sc i} 10830 \AA{} triplet in prominences and filaments, and by \cite{2011A&A...529A..12K, 2012A&A...543A.158K, 2012A&A...545A..52K, 2016FrASS...3...20R} and \cite{2017ApJ...838...69L} in different solar coronal lines. 

\begin{figure}
    \centering
    \includegraphics[scale=0.65]{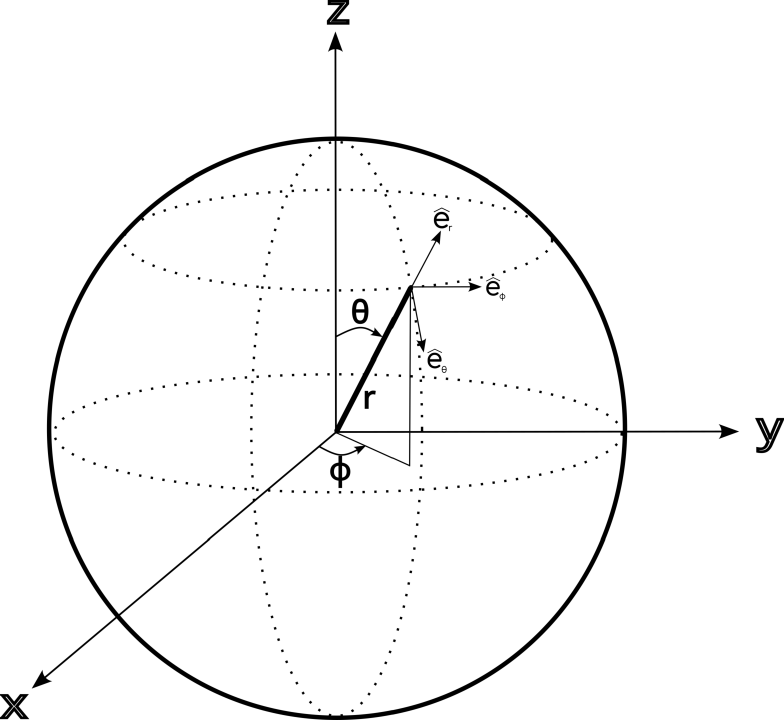}
    \caption{Representation of the coordinate axes in \mbox{P-CORONA}. The X-axis is parallel to the line-of-sight (LOS), and therefore the YZ plane is perpendicular to the LOS. The magnetic and velocity vector field components are given in spherical coordinates with reference to the “local vertical” reference system, defined by the unit vectors \( (\hat{\mathbf{e}}_r, \hat{\mathbf{e}}_\theta, \hat{\mathbf{e}}_\phi) \), at each grid point.}
    \label{coord}
\end{figure}

\begin{figure*}
    \centering
    \includegraphics[scale=0.5]{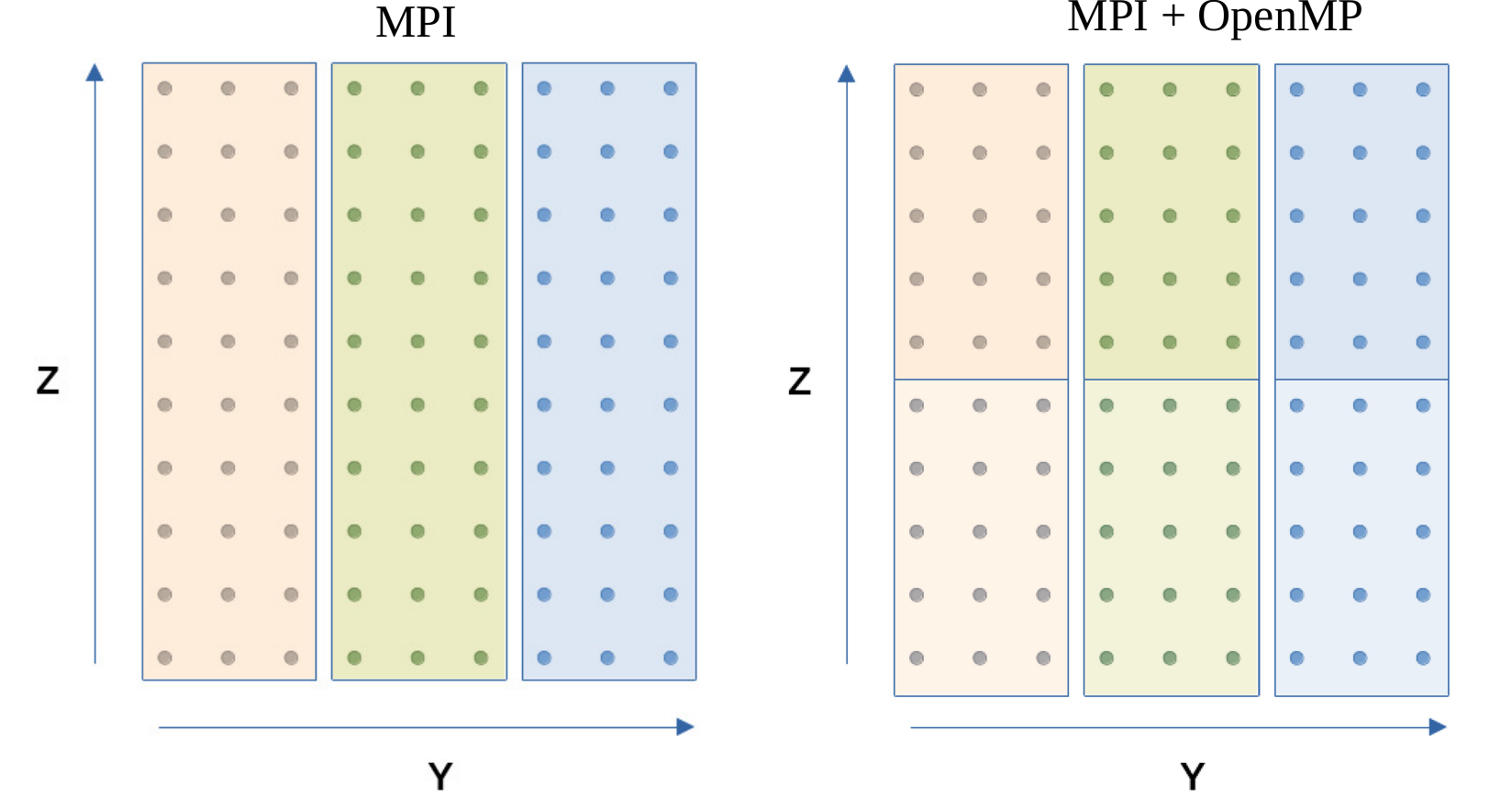}
    \caption{Illustrative example of the spatial domain
        decomposition with MPI and hybrid MPI+OpenMP parallelization in \mbox{P-CORONA}.}
    \label{mpi-openmp}
\end{figure*}

This paper introduces a novel forward modeling code, \mbox{P-CORONA}, which consistently models both the forbidden and permitted lines in three-dimensional (3D) coronal models. The main motivation for developing a new code was to have a single user-friendly tool that can model the polarization of both dipolar forbidden and permitted coronal spectral lines by accounting for the relevant physical mechanisms (see Section~\ref{sec:theoreticalformulation}). \mbox{P-CORONA} is developed following the density matrix theory of spectral line polarization described in \citet[][hereafter, LL04]{LL04}. We believe the application of this code will improve our understanding of coronal structures and dynamics, paving the way for a more accurate understanding of the physical processes influencing the solar corona. 
For the benefit of the community, \mbox{P-CORONA} is made available as an open-source code and can be accessed on Zenodo: \dataset[doi:10.5281/zenodo.15195460]{https://doi.org/10.5281/zenodo.15195460} \citep{hebbur_dayananda_2025_15195461}. The authors have also compiled detailed documentation about the code at \href{https://polmag.gitlab.io/P-CORONA/index.html}{https://polmag.gitlab.io/P-CORONA/index.html}. 
In Section~\ref{sec:framework} we elaborate on the technical aspects of the implementation, 
and in Section~\ref{sec:Results} we show a practical application to highlight the importance and advantage of having a forward modeling tool of this nature. In Section~\ref{sec:zeeman_effect} we elaborate on one of the salient features of \mbox{P-CORONA}, namely the consistent treatment of the Zeeman effect and its signatures in different polarization signals.

\section{P-CORONA: physics and implementation} \label{sec:framework}
The motivation for developing \mbox{P-CORONA} was to create a plasma diagnostic tool that could accurately model both permitted and forbidden spectral lines in 3D models of the solar corona. Existing tools often
specialize in particular cases, such as Hanle-saturated forbidden lines, or lack the capability to handle the full complexity of physical processes affecting these lines —namely, the simultaneous influence of magnetic fields, the Doppler dimming or brightening that modifies the resonant scattering due to bulk radial velocities, and collisional excitation. By accounting for all these physical ingredients, \mbox{P-CORONA} aims to streamline calculations of the Stokes parameters for coronal spectral lines.

The theoretical formulation of \mbox{P-CORONA} is based on the complete frequency redistribution theory of spectral line polarization (see LL04). The steps performed by \mbox{P-CORONA} can be summarized as follows. In the first step, a set of input parameters — the atomic model, the coronal atmospheric model, and the radiation emerging from the underlying atmosphere — is specified for the problem of interest. In the second step, the statistical equilibrium equations (SEE) for the multipolar components of the atomic density matrix elements ($\rho^K_Q$) are solved using the input parameters specified in the first step. 
The resulting density matrix elements quantify the 
population of the magnetic sublevels and the quantum interference among them.
Finally, the emissivities for all Stokes parameters—$I$, $Q$, $U$, and $V$—are computed and then integrated along the LOS to obtain either the spectral variation of the Stokes profiles or the frequency-integrated signal for the line of interest. In this section, we elaborate on the details of each of these steps and also discuss the parallelization strategy of the code.

\subsection{Inputs} \label{sec:inputparameters}
\mbox{P-CORONA} requires three primary sets of input parameters, all specified in the HDF5 format. The first set pertains to the radiation emerging from the underlying solar atmosphere that reaches or illuminates the coronal atoms. For most forbidden lines, typically in the IR spectral range, the relevant solar disk radiation is that of the continuum. However, for permitted lines such as the Ly-$\alpha$ lines of H {\sc i} and He {\sc ii}, the coronal atoms see a strong emission profile from the disk \citep[e.g.,][]{2021ApJ...920..140S}. \mbox{P-CORONA} considers both scenarios. 

The second set of input parameters includes the atomic model for the coronal ion of interest. \mbox{P-CORONA} requires atomic data such as radiative and collisional transition rates, atomic-level energies, and ionization fractions to model the intensity and polarization of spectral lines. These data can be obtained from databases like CHIANTI, from which we have taken the data for our calculations \citep[version 10;][]{2021ApJ...909...38D}. The abundances are obtained from \citet{2012ApJ...755...33S}, and are assumed constant in the solar corona. It is important to note that the choice of elemental abundances is not unique, and the composition of coronal abundances remains an active area of research \citep[e.g.,][]{2018ApJ...852...52D}. 

\begin{figure}
    \centering
    \includegraphics[scale=0.37]{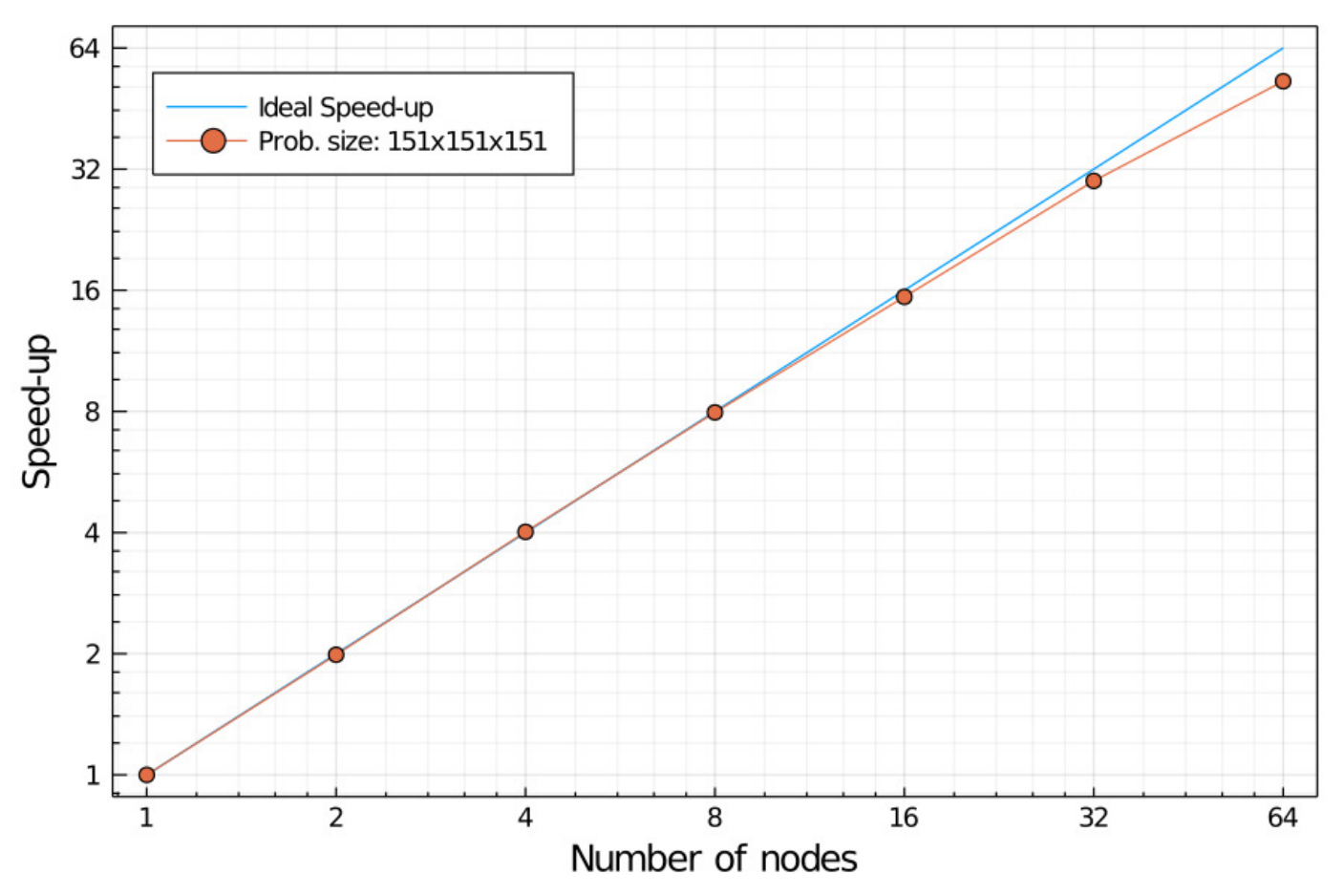}
    \caption{P-CORONA real and ideal speed-up when run with a different number of nodes (with 36 OpenMP threads per node) in the Piz Daint supercomputer. The largest run used 2304 cores.}
    \label{speedup}
\end{figure}

The third set of input parameters is the 3D coronal atmospheric model. This includes physical properties such as the electron temperature and number density, and the magnetic and velocity field vectors at each point in the 3D coronal volume. 
\mbox{P-CORONA} uses a right-handed coordinate system centered at the Sun and a Cartesian grid for the discrete representation of the spatial distribution of physical quantities. The LOS is along the X-axis, and the YZ plane is perpendicular to the LOS as shown in Figure \ref{coord}. This fixed-LOS geometry is suitable for most remote-sensing observations, although it may not fully capture the varying LOS directions encountered in near-Sun missions such as Solar Orbiter or Parker Solar Probe. The magnetic and macroscopic velocity vector field components are specified in spherical coordinates with reference to the ``local vertical" reference system, defined by the unit vectors \( (\hat{\mathbf{e}}_r, \hat{\mathbf{e}}_\theta, \hat{\mathbf{e}}_\phi) \) shown in Figure \ref{coord}, at each grid point. We refer the reader to the documentation of \mbox{P-CORONA}, made available in the public version of the code, for details on the format of the files for each of these input paramet
ers (see \href{https://polmag.gitlab.io/P-CORONA/Appendices/index.html}{https://polmag.gitlab.io/P-CORONA/Appendices/index.html}). 

\begin{figure*}
    \centering
    \includegraphics[scale=0.5]{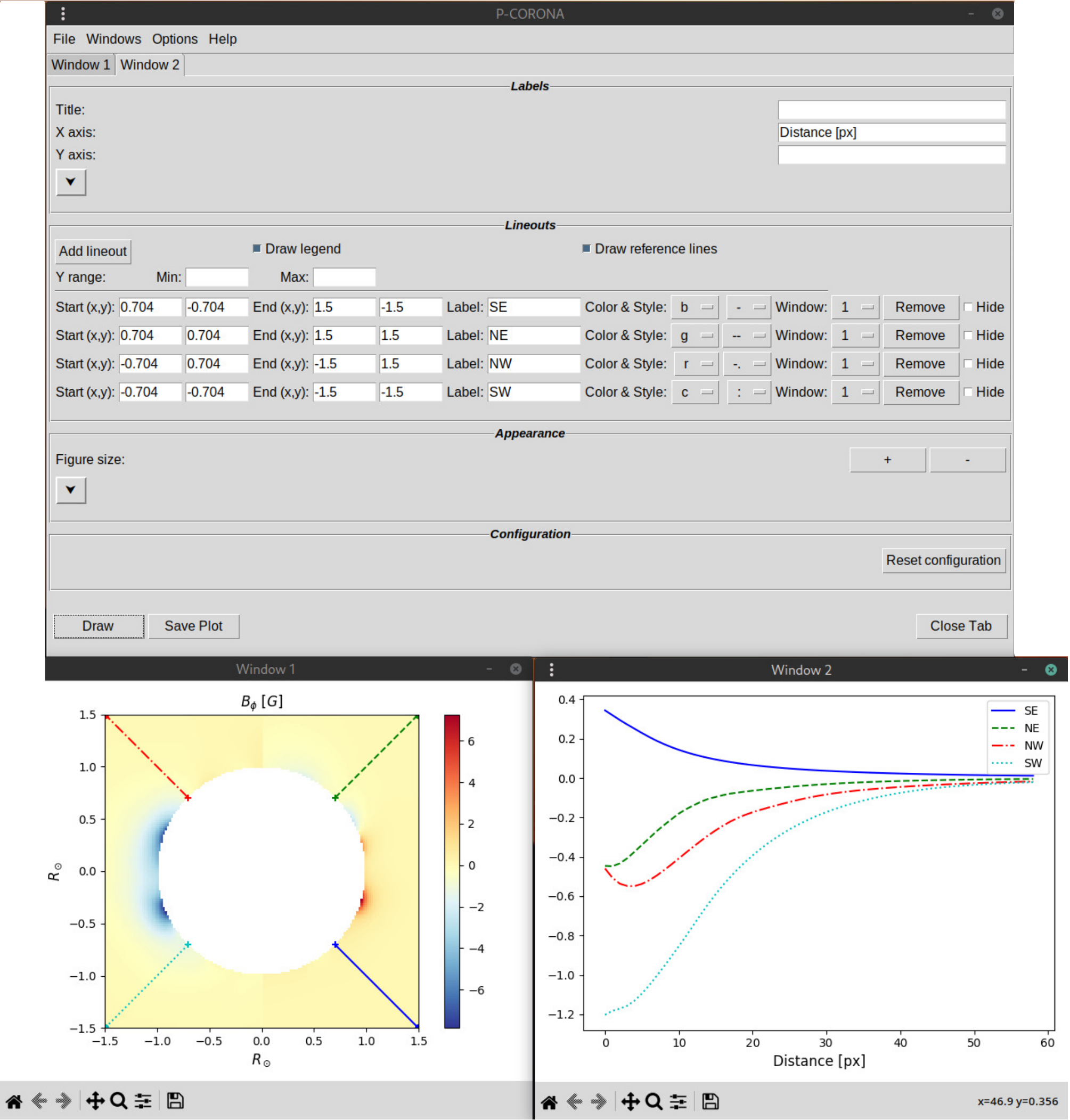}
    \caption{Sample visualization of lineout data (1D variation along a specific direction) using the GUI of \mbox{P-CORONA}.}
    \label{lineout}
\end{figure*}

\subsection{Theoretical formulation} \label{sec:theoreticalformulation}
\mbox{P-CORONA} accounts for various physical mechanisms affecting the intensity and polarization of spectral lines: scattering of anisotropic radiation, collisions with electrons and protons, the impact of magnetic field through the Hanle and Zeeman effects, and Doppler dimming and brightening. In this section, we briefly explain the approach for their implementation.

\mbox{P-CORONA} is based on the density matrix theory of spectral line polarization under complete frequency redistribution, as detailed in Section 7.2 of LL04.  Here we provide some of the relevant equations for completeness and clarity. \mbox{P-CORONA} solves the SEEs for a multi-level atomic system using the spherical tensor representation of the atomic density matrix. This formalism accounts for both radiative and collisional transitions among the atomic levels. The solar radius vector passing through the spatial point of interest is used as the quantization axis for the total angular momentum.

The atomic state is described by the multipolar components $\rho^K_Q(\alpha J)$ of the atomic density matrix, where $J$ represents the total angular momentum quantum number and $\alpha$ additional quantum numbers specifying the level. The evolution of these components is governed by the magnetic field, radiative excitation and de-excitation (via absorption, spontaneous and stimulated emission), and collisional processes. The radiative rates corresponding to absorption and stimulated emission processes depend on the radiation field tensors $J^K_Q$, which quantify the angular distribution and anisotropy of the incident radiation. These tensors are affected by Doppler dimming or brightening due to bulk motions in the corona \citep[see \mbox{Equation (2)} in][ along with the subsequent discussion in that paper]{2021ApJ...920..140S} and, in general, by surface brightness inhomogeneities on the solar disk. While the current version of P-CORONA accounts for the impact of the Doppler dimming or brightening on the emergent Stokes profiles, the inclusion of disk inhomogeneities is planned for a future development. Collisional rates include inelastic, superelastic, and elastic contributions due to electrons and protons.

We refer the reader to Section 7.4 of LL04 for the full expressions of the SEEs and the radiative transfer and relaxation rates involved (see particularly Equations 7.14 a–f, 7.78, and 7.101). The expression for the magnetic kernel is provided in Equation (7.79) of LL04, and an earlier derivation is also available in \citet{1990A&A...235..459L}.
 
We point out that the theory presented in LL04 is suitable under the dipole approximation. Considering this, we apply this theory for all the radiative transitions (electric and magnetic dipole and non-dipole transitions) satisfying the selection rules $\Delta J = 0, \pm 1, 0 \not \rightarrow 0$. Any other radiative transition is accounted for under the strong coupling approximation, as outlined in \citet{2006ApJ...651.1229J} and \citet{2020SoPh..295...98S}. Under this approximation, the transition rates are assumed to contribute only to the population of the atomic levels and not to its polarization. A new formalism proposed by \citet{2024arXiv240901197C} extends the LL04 theory of polarized line formation to include electric and magnetic multipole radiative transitions. In the future, we plan to explore its implementation in \mbox{P-CORONA}.

Concerning the collisional transitions between levels connected by electric dipolar or quadrupolar radiative transitions, as detailed in Appendix A4 of LL04, they are accounted for using the Born approximation. It is described by just one operator of rank $\tilde{K}$ \footnote{$\tilde{K}$=1,2 respectively for dipole and quadrupole transitions}. Any other collisional transition is treated under the strong coupling approximation. 

After solving the SEEs for the multipolar components of the atomic density matrix, $\rho^K_Q (J)$, the emissivity for each Stokes parameter as a function of line frequency $\nu$ and propagation direction $\vec{\Omega}$ is computed as (see equation (7.15e) of LL04) 
\begin{eqnarray}
&& \epsilon_{i}(\nu, \bm {\Omega}) =  \frac{h \nu}{4 \pi} N \sum_{\alpha_{\ell} J_{\ell}} \sum_{\alpha_{u} J_{u}} (2J_{u} + 1) A(\alpha_{u} J_{u} \rightarrow \alpha_{\ell} J_{\ell}) \nonumber \\&&
\quad \times \sum_{K Q K_{u} Q_{u}} \sqrt{3(2K + 1)(2K_{u} + 1)} \nonumber \\&&
\quad \times \sum_{M_{u} M'_{u} M_{\ell} q q'} (-1)^{1+J_{u}-M_{u}+q'} 
\begin{pmatrix}
J_{u} & J_{\ell} & 1 \\
-M_{u} & M_{\ell} & -q
\end{pmatrix} \nonumber \\&&
\quad \times
\begin{pmatrix}
J_{u} & J_{\ell} & 1 \\
-M'_{u} & M_{\ell} & -q'
\end{pmatrix} 
\begin{pmatrix}
1 & 1 & K \\
q & -q' & -Q
\end{pmatrix}
\begin{pmatrix}
J_{u} & J_{u} & K_{u} \\
-M'_{u} & -M_{u} & -Q_{u}
\end{pmatrix} \nonumber \\&&
\quad \times {\rm Re} \left[ T^{K}_{Q}(i, \bm {\Omega}) \rho^{K_{u}}_{Q_{u}}(\alpha_{u} J_{u}) \Phi(\nu_{\alpha_{u} J_{u} M_{u}, \alpha_{\ell} J_{\ell} M_{\ell}} - \nu) \right]. \nonumber \\&&
\label{emissioncoeff}
\end{eqnarray}
Here, the index `{\it i} ' takes the values 0, 1, 2, and 3, corresponding to the Stokes $I$, $Q$, $U$, and $V$ parameters, respectively. $A(\alpha_{u} J_{u} \rightarrow \alpha_{\ell} J_{\ell})$ is the Einstein coefficient for spontaneous emission from the upper ($u$) to the lower ($l$) level of the relevant line transition. The $T^{K}_{Q}$ are the components of the geometrical irreducible spherical tensor depending on the scattering angle and polarization state (see Table 5.6 of LL04). These emissivities incorporate the magnetic quantum numbers, $M$, and the 
spectral line profile functions, $\Phi$, allowing us to account for the impact of the Zeeman effect on both the circular and linear polarization signals in the emergent spectral profiles. The normalized profile is given by
\begin{eqnarray}
 && \Phi(\nu_{\alpha_{u} J_{u} M_{u}, \alpha_{\ell} J_{\ell} M_{\ell}} - \nu) = \phi(\nu_{\alpha_{u} J_{u} M_{u}, \alpha_{\ell} J_{\ell} M_{\ell}} - \nu) 
 \nonumber \\&& \ \ \ \ \ \ \ \ \ \ \ \ \ \ \ \ \ \ \ \ \ \ \ \ \ 
 + i\, \psi(\nu_{\alpha_{u} J_{u} M_{u}, \alpha_{\ell} J_{\ell} M_{\ell}} - \nu),
\label{profilefunct}   
\end{eqnarray}
where $\phi$ and $\psi$ are respectively the Voigt and Faraday-Voigt profiles (see Equation (6.59a) of LL04).
Once the emissivities are obtained, \mbox{P-CORONA} integrates them along the LOS to obtain the Stokes profiles using the following expression:
\begin{equation}
I_i (\bm \Omega) = \int \mathrm{d}\nu\ I_i (\nu, \bm \Omega) = \int_{LOS} \epsilon_i (\nu, \bm \Omega) \, \mathrm{d}s.
\label{Eqn-LOS}
\end{equation}
In \mbox{P-CORONA}, we provide the option of computing both frequency-dependent, $I_i (\nu, \bm \Omega)$, and frequency integrated, $I_i (\bm \Omega)$, Stokes parameters. 
Additionally, if the spectral line of interest is in the Hanle-saturated regime, as is often the case for coronal forbidden lines, \mbox{P-CORONA} can perform the above-mentioned computations under this assumption. In the Hanle-saturated regime, the magnetic field is strong enough to destroy the quantum coherences between magnetic sublevels when the quantization axis of total angular momentum is chosen along the magnetic field vector. This simplifies the statistical equilibrium equations by eliminating the need to solve for the off-diagonal density matrix elements, thus, significantly reducing the total computing time. The results obtained from \mbox{P-CORONA} following the general solution for $\rho^K_Q$'s are described in detail in section \ref{sec:Results}. These results were also benchmarked against those presented in \citet{2020SoPh..295...98S}. We found good agreement between the results obtained using both codes, except when accounting for the impact of the Zeeman effe
ct on the Stokes $Q$ and $U$ profiles for some IR lines. Details of these discrepancies are discussed in Section \ref{sec:zeeman_effect}. 

\begin{figure*}
    \centering
    \includegraphics[scale=0.28]{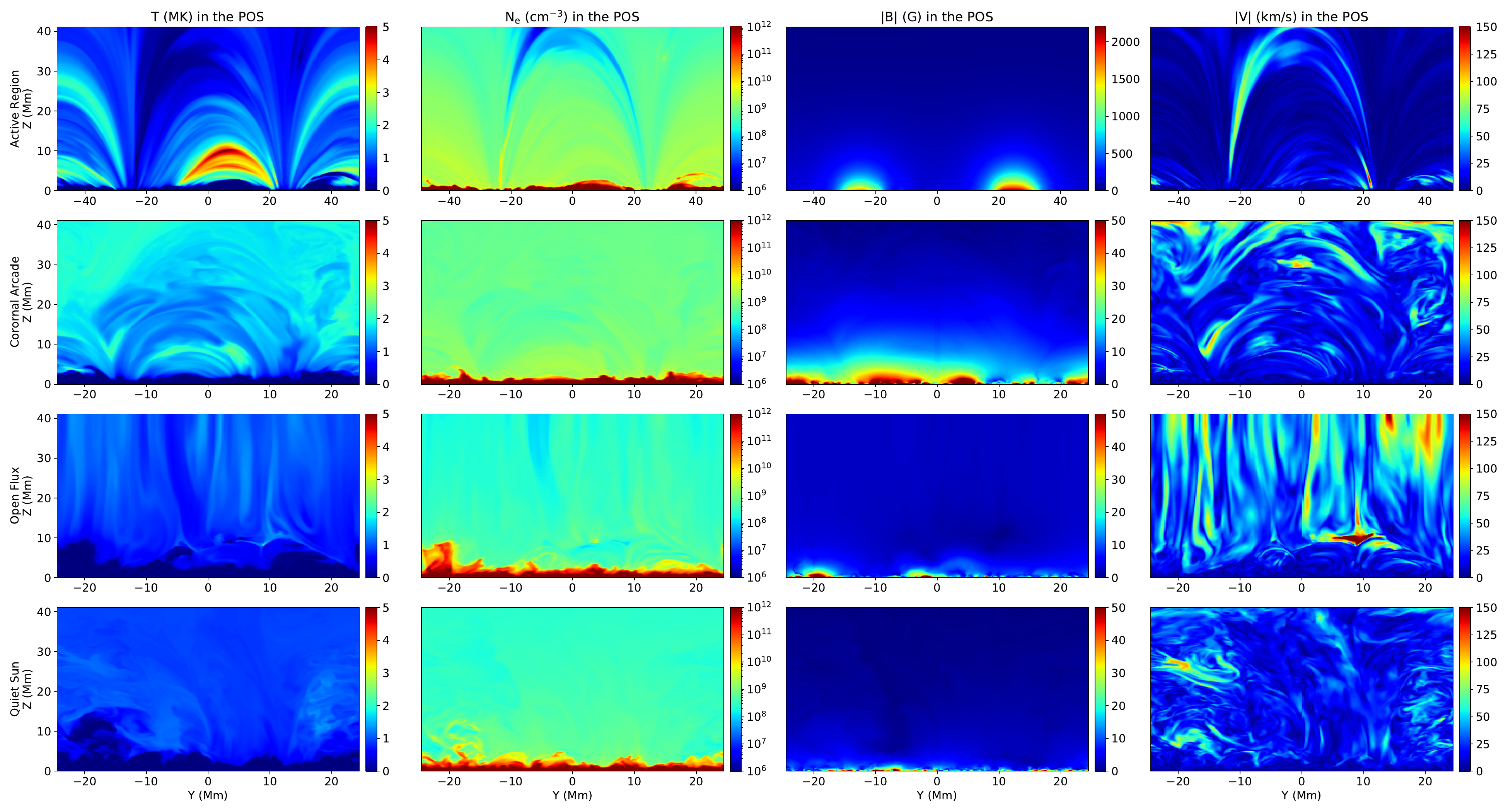}
    \caption{Representation of various atmospheric model parameters from
      different MURaM models, depicted in the plane-of-the-sky. From top to
      bottom, each row illustrates specific quantities corresponding to the Active Region (AR), Coronal Arcade (CA), Open Flux (OF), and Quiet Sun (QS) MURaM models. For each model, the panels are arranged from left to right to show the electron temperature (MK), electron number density (cm$^{-3}$), magnetic field strength (G), and velocity (km/s), respectively.}
    \label{muram_models}
\end{figure*}

\setlength{\tabcolsep}{8pt}
\begin{table*}
    \centering
    \begin{tabular}{c|c|cccc}
        \hline
        {Line} & {$A_{ul}$ (s$^{-1}$)} & 
        \multicolumn{4}{l}{{ I$_{max}$ (ph cm$^{-2}$ s$^{-1}$ arcsec$^{-2}$ nm$^{-1}$)}} \\
         & & {AR} &  {CA} &  {OF} &  {QS}  \\ \hline
        Fe {\sc xiv} 5303 \AA{} & 55.2 & 9897.07 & 7002.22 & 5.07 & 1.99 \\ 
        ($J_u = 3/2 \rightarrow J_l= 1/2$) & & & & \\
        & & & & &  \\
        Fe {\sc xi} 7892 \AA{} & 43.9 & 4218.64 & 1290.40 & 191.30 & 506.43 \\ 
        ($J_u = 1 \rightarrow J_l= 2$) & & & & \\ 
        & & & & & \\
        Fe {\sc xiii} 10747 \AA{} & 14.0 & 2543.35 & 2552.84 & 15.50 & 15.36 \\
        ($J_u = 1 \rightarrow J_l= 0$) & & & & \\
        & & & & & \\
        Fe {\sc xiii} 10798 \AA{} & 9.88 & 2129.52 & 1942.49 & 9.19 & 9.17 \\ 
        ($J_u = 2 \rightarrow J_l= 1$) & & & & \\
        & & & & & \\
        Si {\sc x} 14301 \AA{} & 3.08 & 778.42 & 468.05 & 39.17 & 78.42 \\ 
        ($J_u = 3/2 \rightarrow J_l= 1/2$) & & & & \\
        & & & & & \\
        Si {\sc ix} 39343 \AA{} & 0.30 & 20.17 & 7.22 & 5.32 & 12.37 \\ 
        ($J_u = 1 \rightarrow J_l= 0$) & & & & \\ \hline
    \end{tabular}
      \caption{Maximum intensity for the chosen forbidden lines in each of the MURaM models, as shown in the last row of Figures \ref{fe13_10747} to \ref{si10_14301}. The lines are tabulated in descending order of their Einstein coefficient for spontaneous emission. In the first column, the ion transition wavelength is presented along with the total angular momentum $J$ for the upper and lower levels involved in the transition.}
    \label{tab:stokesI}
\end{table*}

\subsection{Parallelization and GUI in \mbox{P-CORONA}}
\mbox{P-CORONA} is written in Fortran 90 and it is parallelized with MPI and OpenMP. It can be run in parallel using either pure MPI or hybrid MPI+OpenMP parallelization. For pure MPI parallelization, the spatial domain in the Y-direction is divided into $n - 1$ blocks ($n$ being the total number of MPI processes used, since one process is needed for coordination and I/O). 
Figure~\ref{mpi-openmp} illustrates the case where four MPI processes are used, with the Y-direction of the domain divided accordingly among them. 
For the hybrid MPI+OpenMP parallelization, in addition to the division
in the Y-direction, we further subdivide the spatial domain
in the vertical Z-direction into a number of OpenMP threads. For example, if we request to use four MPI processes and two OpenMP threads then the three blocks in the Y-direction are further divided into two blocks in the Z-direction, as illustrated in the right panel of Figure~\ref{mpi-openmp}. This hybrid approach results in a better performance and smaller memory footprint of \mbox{P-CORONA} when run in a multi-node setting. In this scenario, the recommended approach is to run one MPI process per node, while using as many OpenMP threads as CPU cores there are in each node. In Figure~\ref{speedup} we show the parallelization speed-up of \mbox{P-CORONA} with the hybrid approach for a medium-size spatial domain (151 x 151 x 151 grid points) run in the Piz Daint supercomputer, always with 36 OpenMP processes (a full node), for an increasing number of nodes. Only for the largest number of cores considered (64 nodes for a total of 2304 cores) we get a speed-up appreciably below t
he ideal, showing the excellent strong scaling of \mbox{P-CORONA}. 
Information on the total computing time required by \mbox{P-CORONA} for forward modeling with 
increasingly complex atomic levels of Fe {\sc xiii} can be found in \citet{delzanna25}.

\mbox{P-CORONA} also provides a graphical user interface (GUI) to help manage and visualize the input and output files. This tool can generate 2D images of coronal atmospheric quantities and frequency-integrated Stokes
parameters, draw line-outs within these 2D images to study the variation of
different quantities along a specific direction and plot the spectral variation
of the Stokes profiles. A sample visualization of lineout data using the GUI is shown in Figure~\ref{lineout}. Extensive details on this are provided in the code documentation at \href{https://polmag.gitlab.io/P-CORONA/Visualization/index.html}{https://polmag.gitlab.io/P-CORONA/Visualization/index.html}.

\setlength{\tabcolsep}{10pt}
\begin{table*}
\centering
    \begin{tabular}{c|cccc|cccc}
\hline
        {Line} &  
        \multicolumn{4}{l|}{{ P$_{max}$ ($\times$ 10$^{-3}$ ph cm$^{-2}$ s$^{-1}$ arcsec$^{-2}$ nm$^{-1}$)}} &
        \multicolumn{4}{l}{{ ${\bf |\rm V_{max}|}$ ($\times$ 10$^{-3}$ ph cm$^{-2}$ s$^{-1}$ arcsec$^{-2}$ nm$^{-1}$)}} \\
& {AR} & {CA} & {OF} & {QS} & 
             {AR} & {CA} & {OF} & {QS} \\ \hline
        Fe {\sc xiv} 5303 \AA{} & 
               4668.15 & 3826.90 & 20.09 & 1.32 &
               1283.80 & 368.77 & 0.10 & 0.002 \\ 
        ($J_u = 3/2 \rightarrow J_l= 1/2$)& & & & & & & &\\ 
        & & & & & & & \\
        Fe {\sc xi} 7892 \AA{} &  
               99.43 & 0.69 & 6.17 & 4.0 & 
               479.34 & 99.99 & 2.36 & 6.76 \\ 
        ($J_u = 1 \rightarrow J_l= 2$)& & & & & & & & \\ 
        & & & & & & & \\
        Fe {\sc xiii} 10747 \AA{} & 
        2809.35 & 3883.17 & 185.65 & 35.69 &
               315.57 & 260.23 & 0.45 & 0.29 \\
        ($J_u = 1 \rightarrow J_l= 0$)& & & & & & & & \\
        & & & & & & & \\
        Fe {\sc xiii} 10798 \AA{} &   
               374.65 & 367.81 & 13.86 & 2.86 & 
               263.18 & 201.0 & 0.26 & 0.17 \\ 
        ($J_u = 2 \rightarrow J_l= 1$)& & & & & & & & \\
        & & & & & & & \\
        Si {\sc x} 14301 \AA{} &   
               317.93 & 271.77 & 185.15 & 100.82 & 
               69.31 & 49.02 & 6.14 & 1.09 \\ 
        ($J_u = 3/2 \rightarrow J_l= 1/2$)& & & & & & & & \\
        & & & & & & & \\
        Si {\sc ix} 39343 \AA{} &   
               5.55 & 0.08 & 2.34 & 1.36 & 
               11.12 & 2.20 & 0.21 & 0.71\\ 
        ($J_u = 1 \rightarrow J_l= 0$)& & & & & & & & \\ \hline
    \end{tabular}
      \caption{Same as Table \ref{tab:stokesI} but tabulating the 
      total linear polarization and an absolute maximum of the circular polarization for the chosen forbidden lines in each of the MURaM models, as shown in the last row of Figures \ref{fe13_10747} to \ref{si10_14301}.}
    \label{tab:stokesPV}
\end{table*}

\section{Application} \label{sec:Results}
With \mbox{P-CORONA} we can investigate the intensity and polarization of coronal spectral lines, from extreme ultraviolet (EUV) to IR wavelengths, to understand the behavior of the corresponding Stokes parameters in 3D models of the solar corona. 
In this section, we consider a few illustrative examples to demonstrate the
capabilities of \mbox{P-CORONA}. We focus on the coronal forbidden lines. The
capabilities of \mbox{P-CORONA} for permitted lines like the Ly-$\alpha$ lines of H
{\sc i} and He {\sc ii} were previously demonstrated in
\citet{2021ApJ...920..140S}. For the computations presented here, we consider
two sets of models to illustrate the behavior of the synthesized polarization
signals for different forbidden spectral lines. The 3D models include numerical
simulations from MURaM extended to coronal heights \citep{2017ApJ...834...10R}
and from Predictive Science Inc. (PSI) models (see
\href{https://www.predsci.com/portal/home.php}{https://www.predsci.com/portal/home.php}). Choosing these two types of 3D coronal models helps demonstrate the ability of \mbox{P-CORONA} in modeling spectral lines in both a 3D coronal model (as in MURaM) developed for lower coronal heights and in a large spherical 3D coronal simulation (as in the PSI models), which extends up to several solar radii.

\begin{figure*}
    \centering
    \includegraphics[scale=0.3]{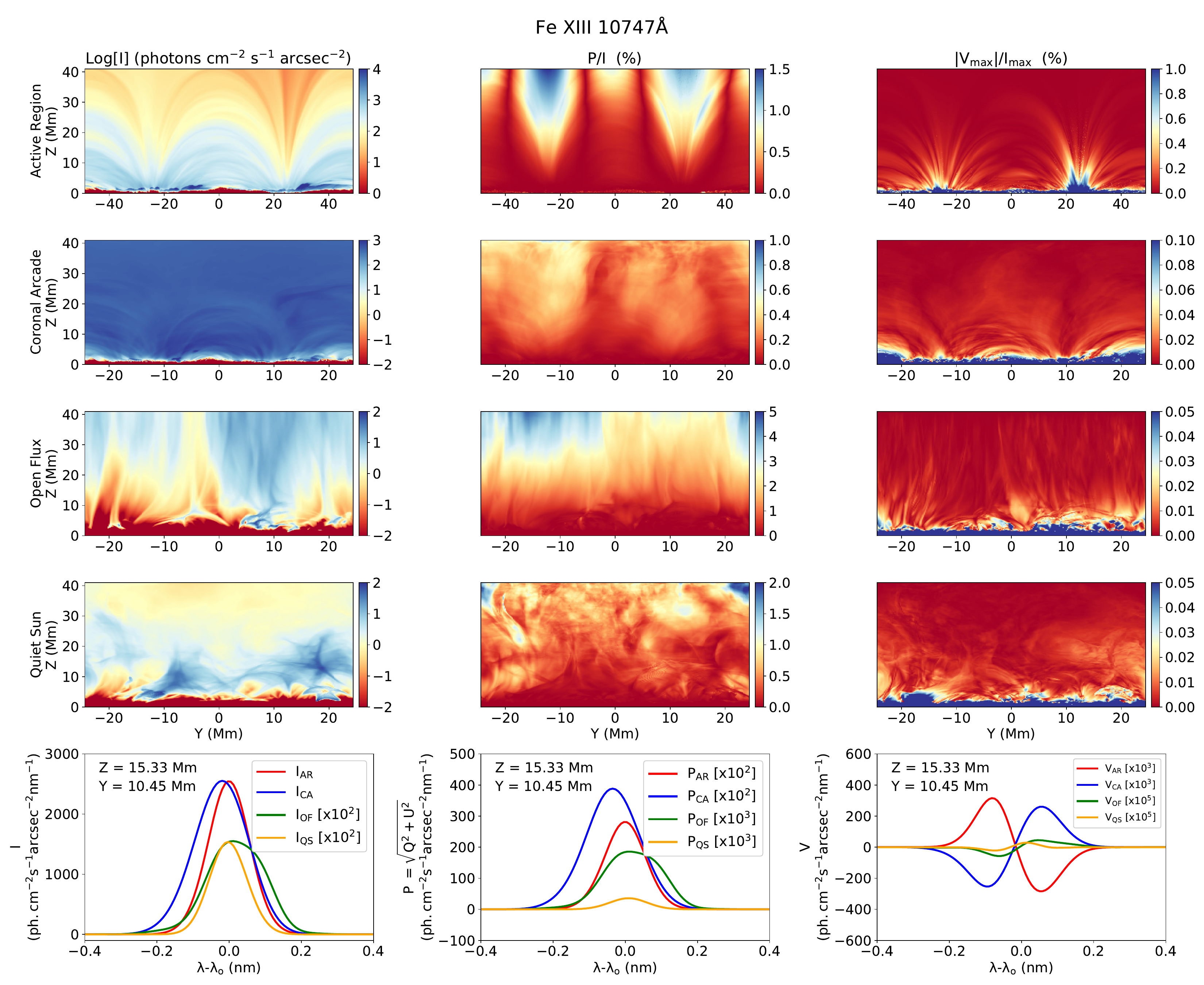}
    \caption{Stokes parameters for the Fe {\sc xiii} 10747 \AA{} line in selected MURaM models. Each of the first four rows presents maps of the frequency-integrated intensity (column 1), the total fractional linear polarization (column 2), and the maximum of Stokes V relative to the intensity I (column 3) for different MURaM models. The last row shows, from left to right, the intensity, total linear polarization, and circular polarization profiles, respectively, at a specific point (Y=10.45 Mm and Z=15.33 Mm) in the MURaM models. {Note that differences in the colorbars across panels.}}
    \label{fe13_10747}
\end{figure*}

\subsection{The polarization of forbidden lines in MURaM Models} \label{FS-muram}
The MURaM models of the lower corona are detailed in \citet{2017ApJ...834...10R}. We selected four MURaM snapshots, each representing different coronal conditions: Active Region (AR), Coronal Arcade (CA), Open Flux (OF), and Quiet Sun (QS). 
In particular, for these four MURaM coronal models, we examine the behavior of multiple spectral lines under various coronal conditions. Through these studies, we aim to identify the most useful spectral lines for investigating coronal dynamics and magnetism, thereby guiding future observations. Using these models, we specifically calculated the intensity and polarization signals in six forbidden coronal lines: \mbox{Fe {\sc xiii} 10747 \AA{}}, \mbox{Fe {\sc xiii} 10798 \AA{}}, \mbox{Fe {\sc xiv} 5303 \AA{}}, \mbox{Fe {\sc xi} 7892 \AA{}}, \mbox{Si {\sc ix} 39343 \AA{}}, and \mbox{Si {\sc x} 14301 \AA{}}. These lines are particularly important as they are being considered by DKIST for coronal studies \citep{2020SoPh..295...98S}. 

\begin{figure*}
    \centering
    \includegraphics[scale=0.3]{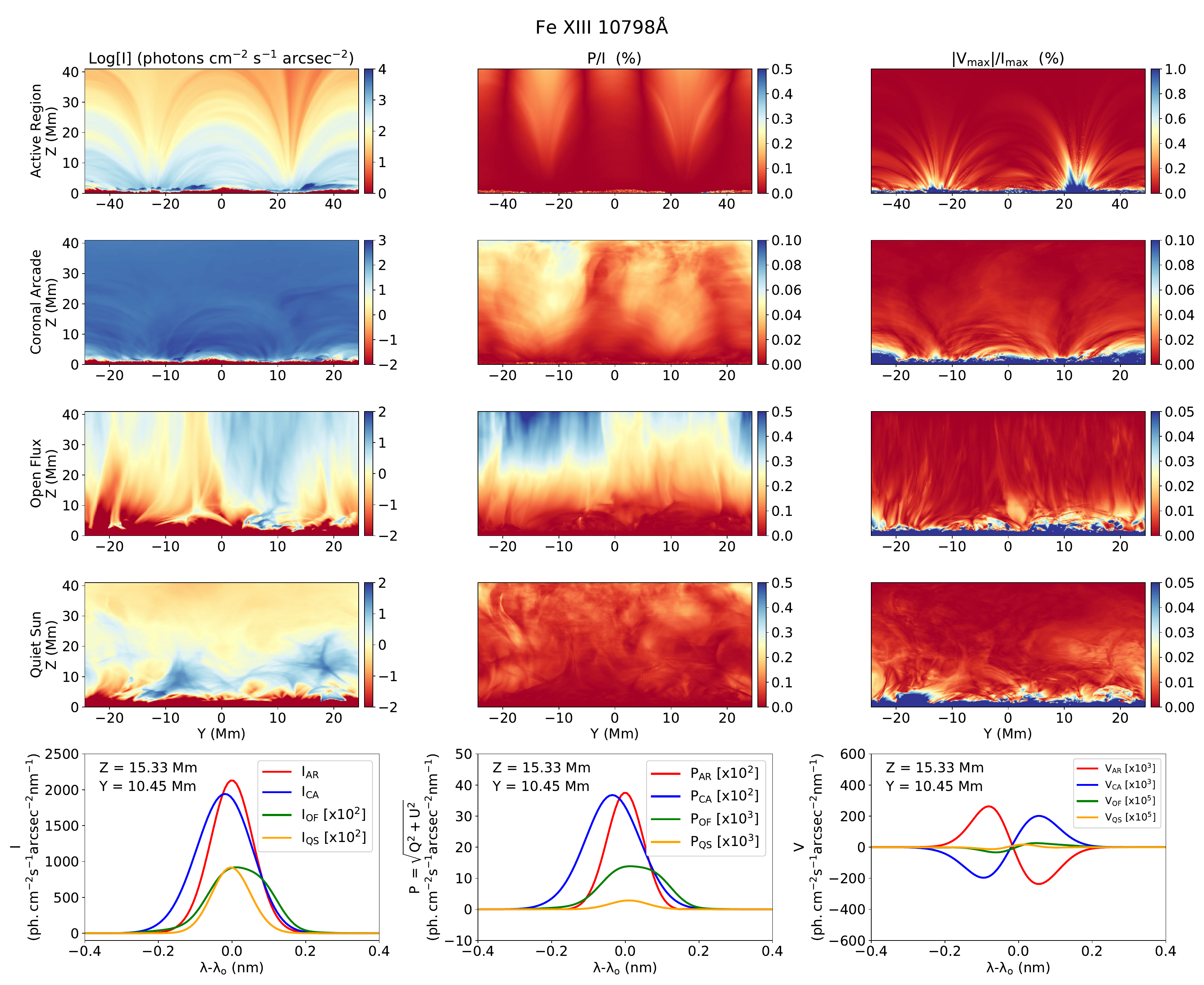}
    \caption{Same as Figure~\ref{fe13_10747}, but for the Fe {\sc xiii} 10798 \AA{} line.}
    \label{fe13_10798}
\end{figure*}

Figure \ref{muram_models} shows the variation of several coronal atmospheric
parameters across the plane of the sky (POS; X=0) in the selected models. 
Each model spans from -24.34 to +24.34 Mm along the LOS (X-axis), and up to ~41 Mm above the photosphere along the Z-axis (see Figure \ref{coord}). Horizontally, along the Y-axis, the models cover -24.34 to +24.34 Mm, except for the AR model, which extends from -48.9 to +48.9 Mm. Figure \ref{muram_models} highlights the significant variation in the physical parameters across the coronal models, depending on the solar coronal structure. For instance, in the POS, the temperature in the OF and QS models reaches approximately 1 MK to 1.5 MK, while in the AR and CA models it reaches between 2 MK and 5 MK. For further details on these models, we refer the reader to the original paper by  \citet{2017ApJ...834...10R}.

\begin{figure*}
    \centering
    \includegraphics[scale=0.3]{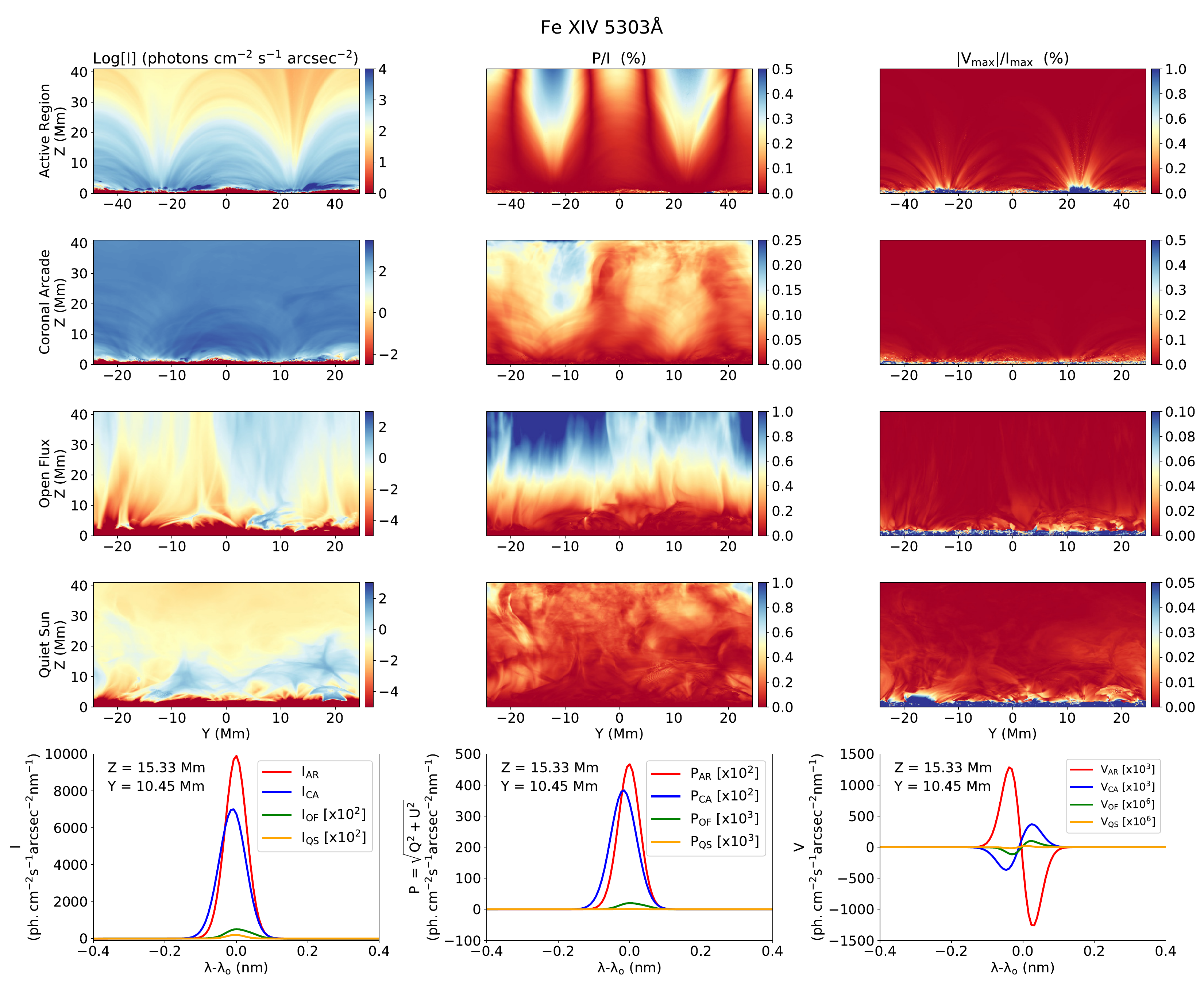}
    \caption{Same as Figure~\ref{fe13_10747}, but for the Fe {\sc xiv} 5303 \AA{} line.}
    \label{fe14_5303}
\end{figure*}

The calculated intensity and polarization signals for the considered set of MURaM models are illustrated in Figures \ref{fe13_10747} to \ref{si10_14301}, respectively for the lines Fe {\sc xiii} 10747 \AA, Fe {\sc xiii} 10798 \AA, Fe {\sc xiv} 5303 \AA, Fe {\sc xi} 7892 \AA, Si {\sc ix} 39343 \AA, and Si {\sc x} 14301 \AA. In each figure, the first four rows show the frequency-integrated signals for the AR, CA, OF, and QS models, respectively. The final row showcases the frequency-dependent LOS integrated signals of each of the Stokes parameters at a specified point in the solar coronal model. As seen in the last row of Figures \ref{fe13_10747} to \ref{si10_14301}, the frequency-dependent intensity and polarization profiles vary significantly depending on the solar conditions.

We list the maximum intensity (Table \ref{tab:stokesI}) and the total linear and circular polarization signals (Table \ref{tab:stokesPV}) for the
  different spectral lines, as observed in the last row of Figures
\ref{fe13_10747} to \ref{si10_14301}. For example, comparing the emission of the
Fe {\sc xiii} 10747 \AA{} line in different models in Table \ref{tab:stokesI} shows
that the maximum intensity at the selected height is similar in both the AR and
CA models. However, the total linear polarization signal in Table
\ref{tab:stokesPV} is stronger in the CA model than in the AR model. Similar
comparisons for other lines reveal that the AR model consistently shows higher
emission in both intensity and polarization compared to the CA
model. Furthermore, within the AR model, the intensity of the different lines at
the selected height directly correlates with the radiative transition
coefficient, with higher intensities observed for lines with larger radiative
transition coefficients. However, this pattern is less consistent in other
models, highlighting the complex interaction between radiative and collisional
transitions, which are influenced by the electron temperature and density in
each model. Comparing the intensity and polarization signals of the selected
lines in the MURaM models reveals that the polarization signals are notably
lower at the lower coronal heights represented in these models, especially in
the OF and QS models. Except for the Fe {\sc xiv} 5303 \AA{} and Fe {\sc xiii} 10747 \AA{}
lines, the measurement of these polarization signals would demand very large
integration times. The high electron density and low anisotropy at these heights
reduce the linear polarization signals. Recent observations of the intensity and
polarization signals of the Fe {\sc xiii} 10747 \AA{} line with DKIST were conducted
in the height range of approximately 1.05 to 1.15 R$_{\odot}$
\citep[see][]{schad_etal:2024}. In the following section, we present the results from
\mbox{P-CORONA} for PSI models extending up to 3 R$_{\odot}$.

\subsection{The polarization of forbidden lines in Predictive Science Models} \label{FS-ps}
This section demonstrates the capabilities of {\mbox{P-CORONA} to handle large atmospheric models, extending up to several solar radii. We analyze two 3D magnetohydrodynamic models of the solar corona and inner heliosphere, CR2157 and CR2138, developed by PSI. A detailed description of these models can be found in \citet{2021ApJ...920..140S}. CR2157, referred to as the `magnetic model', is characterized by a stronger near-limb magnetic activity, while CR2138, referred to as the `dynamic model,' exhibits more prominent macroscopic velocities. Figure~\ref{fe13_10747_PS} shows the calculated frequency-integrated Stokes signals of the Fe {\sc xiii} 10747 \AA{} line in these models. All necessary input parameters to model this line, as well as the other spectral lines discussed in Section~\ref{FS-muram}, are available on Zenodo:
\dataset[doi:10.5281/zenodo.7698835]{https://doi.org/10.5281/zenodo.7698835} along with the open-source version of \mbox{P-CORONA} \citep{hebbur_dayananda_2025_15195461}. 
Using this approach, the behavior of the Stokes parameters in different Predictive Science models for other spectral lines can be explored, allowing for future comparisons similar to those presented in Tables \ref{tab:stokesI} and \ref{tab:stokesPV}.

\section{The Impact of the Zeeman Effect} \label{sec:zeeman_effect}
As described in Section~\ref{sec:theoreticalformulation}, \mbox{P-CORONA} can model the scattering polarization and the Hanle and Zeeman effects in coronal lines. When a magnetic field is inclined with respect to the symmetry axis of the incident radiation field, the Hanle effect modifies the atomic-level polarization \citep[e.g.,][]{2001ASPC..236..161T}. The resulting spectral line polarization is sensitive to magnetic field strengths approximately ranging from 0.2 B$_H$ to 5 B$_H$. Here, B$_H$ is the critical Hanle field intensity (in gauss) for which the Zeeman splitting of the line's level under consideration equals its natural width. This is given by $\rm{B}_{\it H} = 1.137 \times 10^{-7}/(t_{\rm life} g_J)$, where t$_{\rm life}$ is the lifetime of the $J$-level (in seconds), and g$_{\rm J}$ is the level's Lande factor. 
When the magnetic field exceeds \mbox{5 B$_H$}, the Hanle effect approaches saturation, meaning that the linear polarization signals lose sensitivity to the field strength and only depend on its orientation. For forbidden lines, B$_H$ is particularly small due to the long radiative lifetimes of the line's levels, causing Hanle-effect saturation even for very weak magnetic fields.

Even though the Hanle effect becomes insensitive to the magnetic field strength in the saturation regime, the Zeeman effect remains a direct probe of $|B|$. The Stokes $V$ signals are sensitive to the longitudinal component of the magnetic field and are proportional to the ratio $R$ between the Zeeman splitting and the Doppler width. The Stokes $Q$ and $U$ signals, which are sensitive to the transverse component of the magnetic field, scale with $R^2$. Previous modeling efforts of Zeeman signatures in Stokes $V$ \citep{2020SoPh..295...98S,schad_etal:2024} have often used the weak field approximation adjusting for atomic alignment, but the impact of the Zeeman effect on Stokes $Q$ and $U$ has generally been neglected (because of the larger Doppler widths of the coronal lines). \mbox{P-CORONA} computes the emission profiles for Stokes $Q$, $U$, and $V$ using the general expression for emissivity, which fully accounts for magnetic splitting and the density matrix as in Equation 
\ref{emissioncoeff}.
This approach accounts for the impact of the Zeeman effects on all Stokes parameters and not just on Stokes $V$. To show the impact of the Zeeman effect due to the transverse component of the magnetic field in forbidden lines, we examine the \mbox{Fe {\sc xiii} 10747 \AA{}} and the Si {\sc ix} 39343 \AA{} lines. The Zeeman splitting scales with the square of the wavelength, making the Si {\sc ix} 39343 \AA{} line significantly more sensitive to the Zeeman effect than the relatively shorter-wavelength Fe {\sc xiii} 10747 \AA{} line. We now analyze the spectral profiles generated in the AR MURaM model, chosen for its strong magnetic fields. Figures~\ref{zeeman-effectQV} and \ref{zeeman-effectUV} compare the spectral variations of the Stokes parameters at three altitudes in the AR model, with and without the Zeeman effect (solid and dashed lines, respectively). Figure~\ref{zeeman-effectQV} shows the comparison of Stokes $I$, $Q$ and $V$, while Figure~\ref{zeeman-effectUV} compar
es Stokes $I$, $U$ and $V$. In the third column of these Figures, we show the variation of the transverse and longitudinal components of the magnetic field at each of the chosen points. 
The impact of the Zeeman effect due to the transverse component of the magnetic field is present in the computed Stokes $Q$ and $U$ signals, particularly for the Si {\sc ix} 39343 \AA{} line in regions of strong magnetic field (top row, middle panel). For the Fe {\sc xiii} 10747 \AA{} line, however, the corresponding linear polarization signals induced by the transverse Zeeman effect are extremely weak, typically well below 10$^{-4}$ of the peak intensity, and are practically undetectable with current instrumentation. As the altitude in the solar coronal model increases and the model's magnetic field strength weakens, the Zeeman effect due to the transverse component of the magnetic field becomes less noticeable in the Fe {\sc xiii} line and, although still visible in the Si {\sc ix} line, its contribution remains minor.
Nevertheless, even in the most favorable modeled conditions, the linear polarization signals remain at the margins of detectability. Consequently, the diagnostic utility of the transverse Zeeman effect in these forbidden coronal lines is limited. Modest improvements in stray light suppression and longer integration times may enhance future sensitivity, particularly for the Si {\sc ix} line, but the observational challenge remains significant.

\begin{figure*}
    \centering
    \includegraphics[scale=0.3]{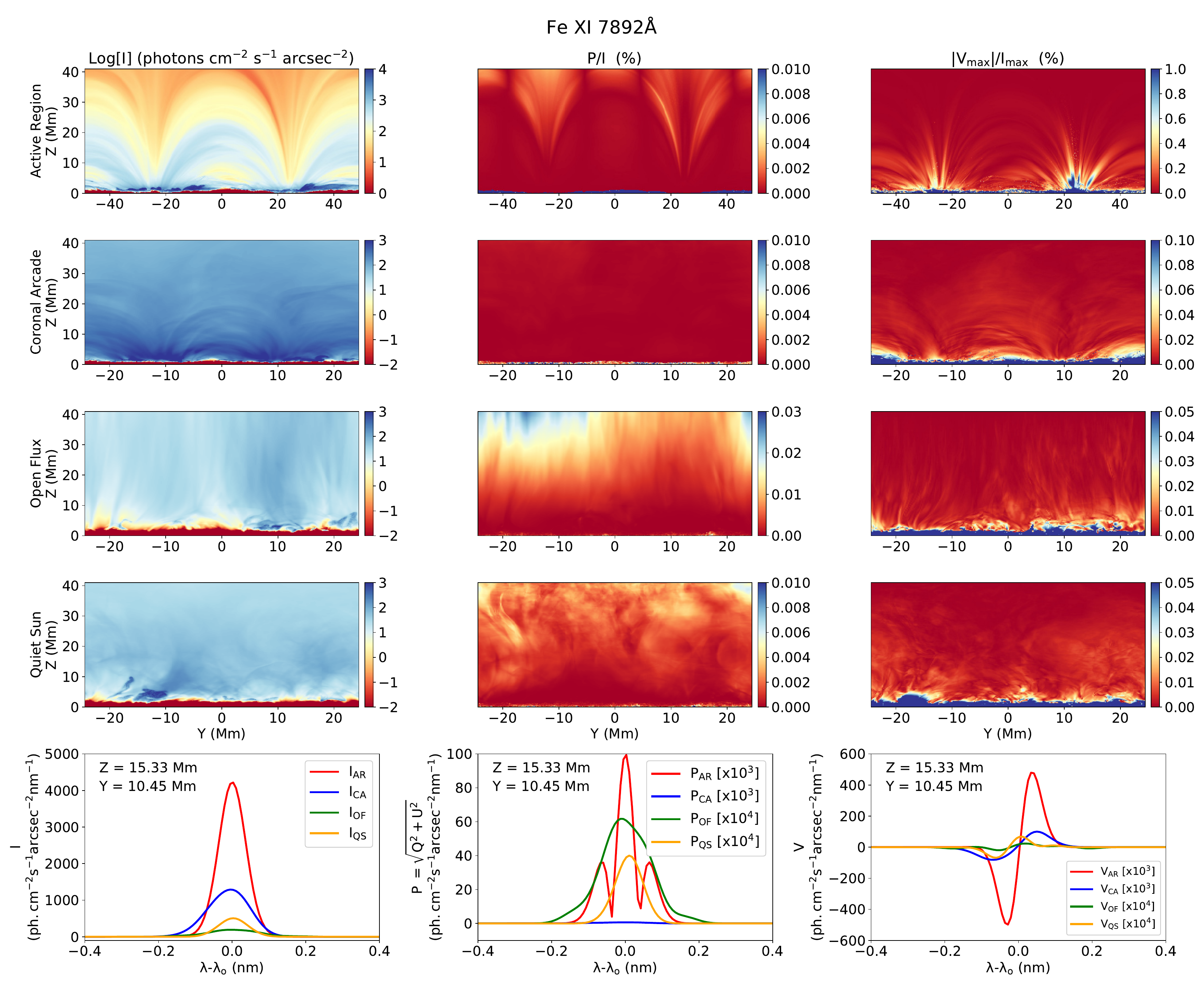}
    \caption{Same as Figure~\ref{fe13_10747}, but for the Fe {\sc xi} 7892 \AA{} line.}
    \label{fe111_7892}
\end{figure*}

\begin{figure*}
    \centering
    \includegraphics[scale=0.3]{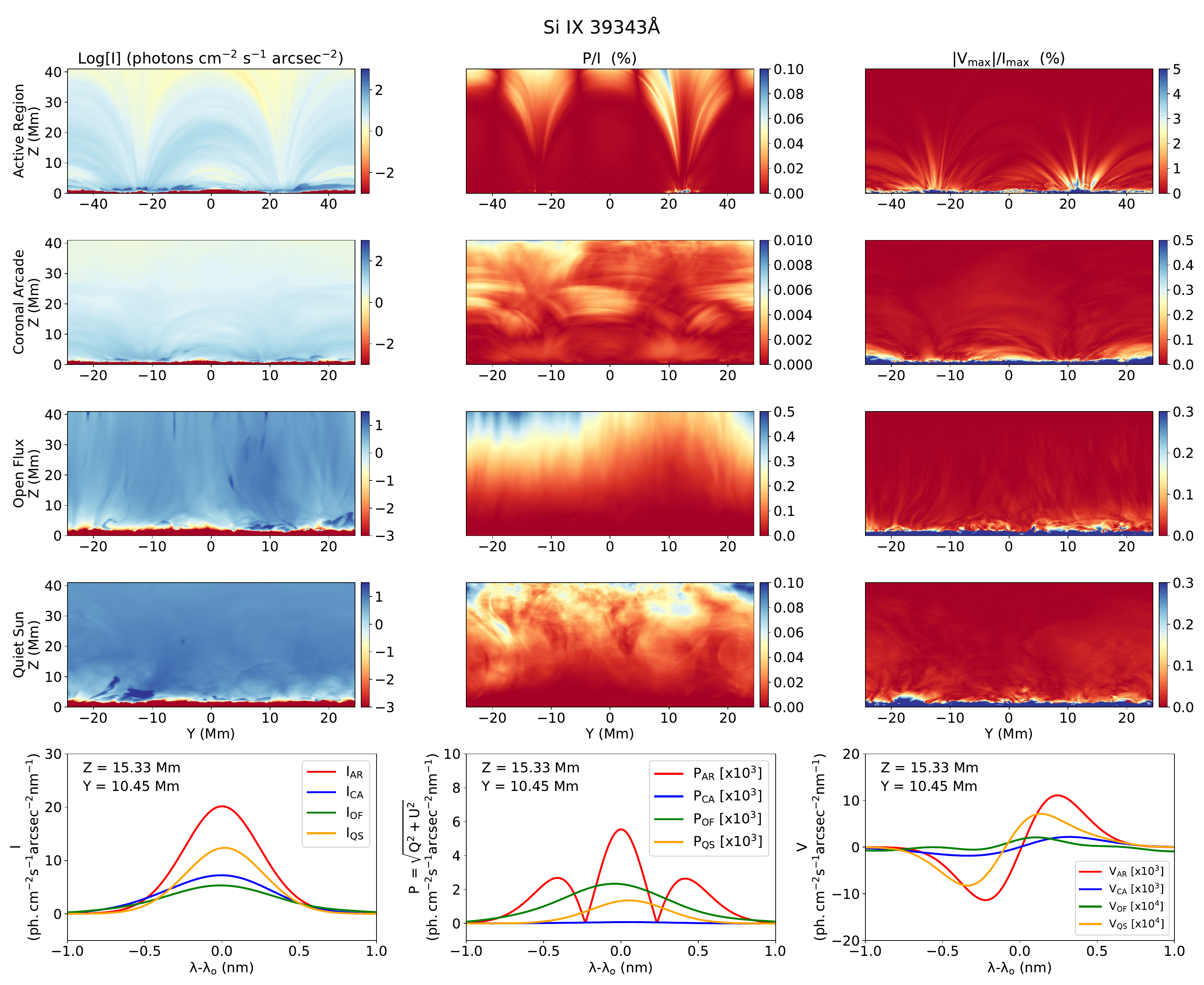}
    \caption{Same as Figure~\ref{fe13_10747}, but for the Si {\sc ix} 39343 \AA{} line.}
    \label{si9_39343}
\end{figure*}

\begin{figure*}
    \centering
    \includegraphics[scale=0.3]{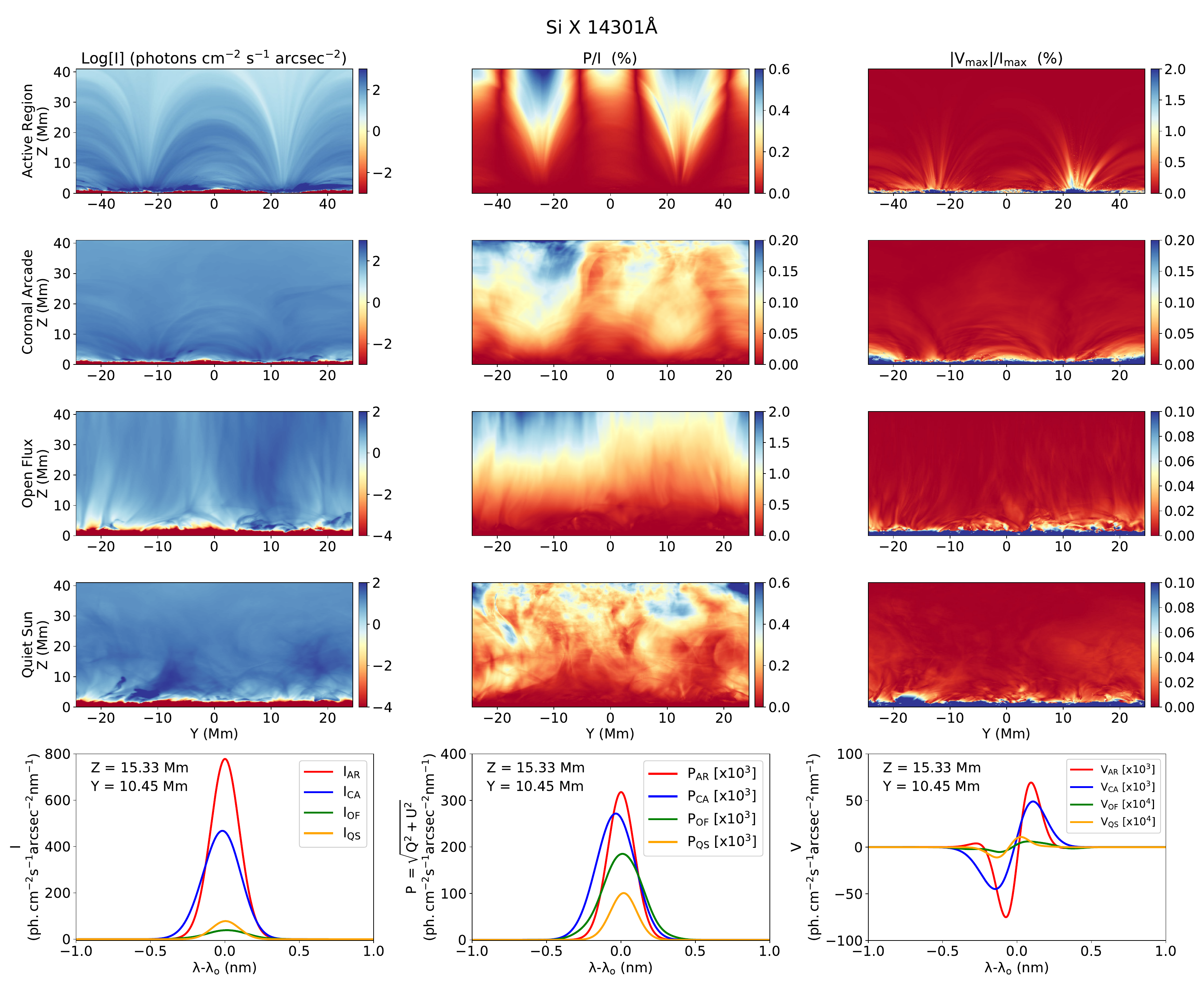}
    \caption{Same as Figure~\ref{fe13_10747}, but for the Si {\sc x} 14301 \AA{} line.}
    \label{si10_14301}
\end{figure*}

\begin{figure*}
    \centering
    \includegraphics[scale=0.25]{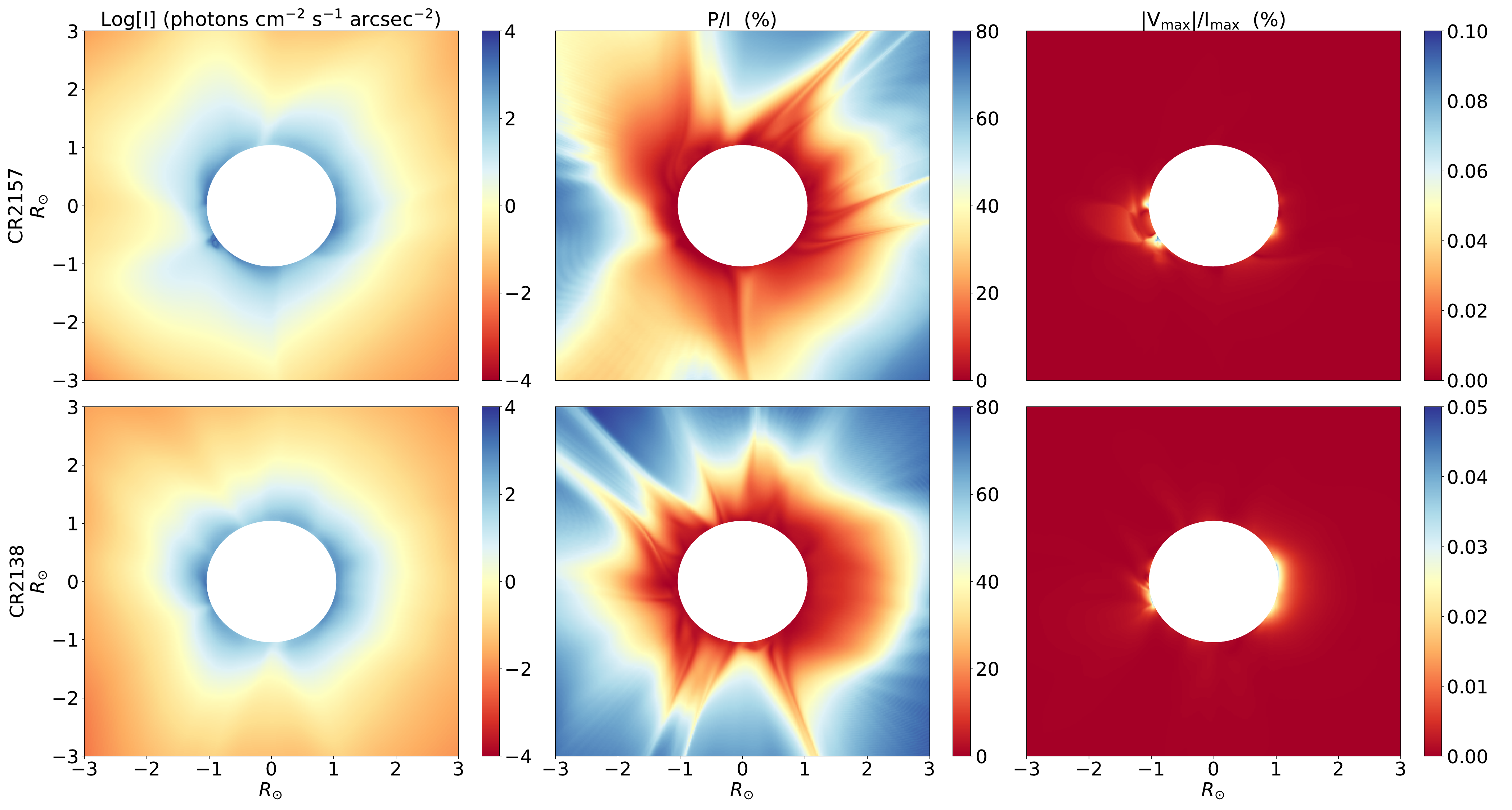}
    \caption{Frequency integrated intensity (left column), total fractional linear
polarization (middle column), and maximum circular polarization relative to the intensity (right column) for the Fe {\sc xiii} 10747 \AA{} line in the Predictive Science models CR2157 (top row) and CR2138 (bottom row).}
    \label{fe13_10747_PS}
\end{figure*}

\begin{figure*}
    \centering
    \includegraphics[scale=0.28]{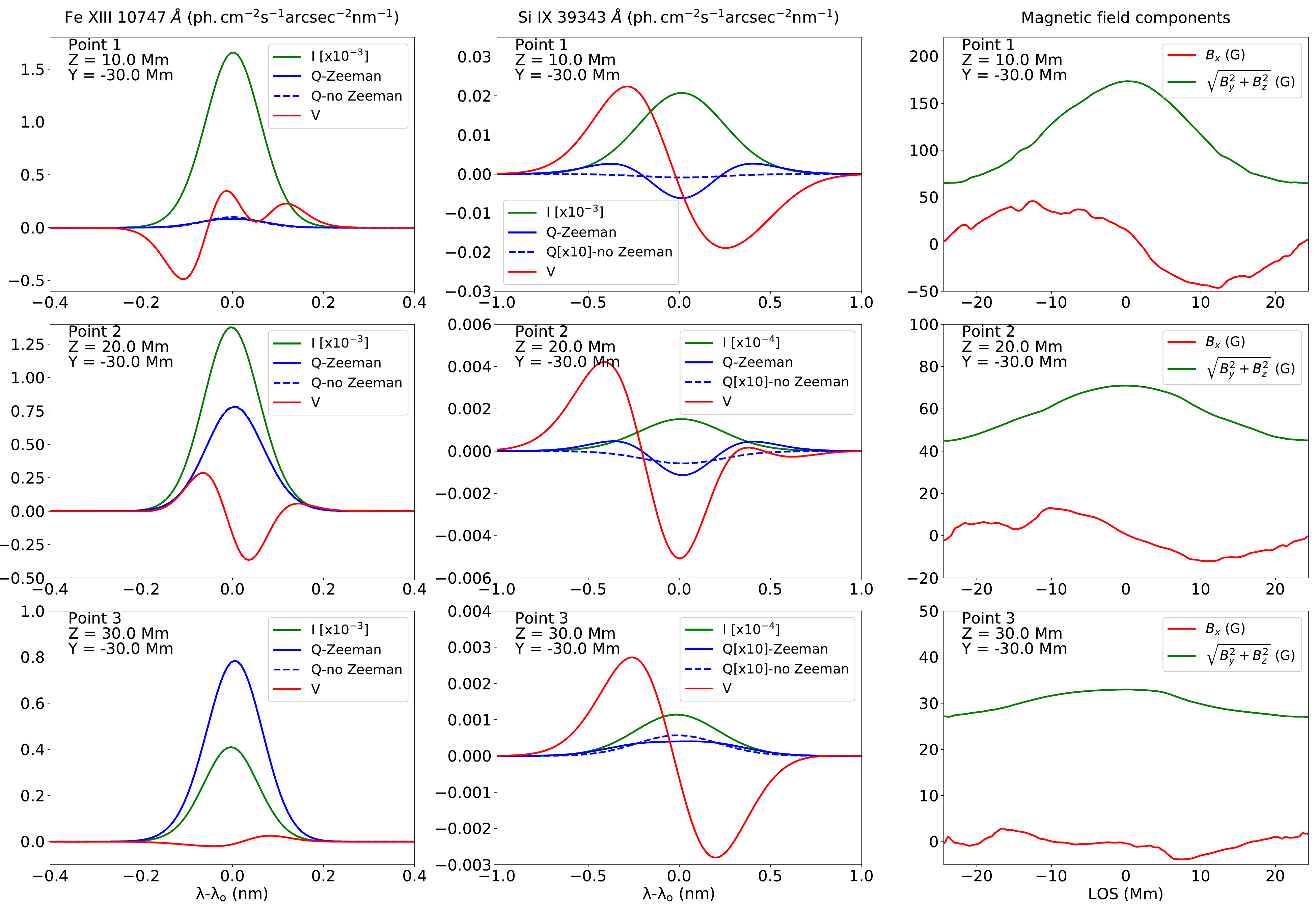}
    \caption{Stokes $\rm I$ (green line), Stokes $\rm Q$ neglecting (black dashed curve) and accounting (solid blue curve) for the Zeeman effect and Stokes $\rm V$ (red curve) profiles for the Fe {\sc xiii} 10747 \AA{} (left column) and Si {\sc ix} 39343 \AA{} (middle column) lines in the AR MURaM model for three different points with coordinates Y=-30 Mm and Z = 10 Mm (top row), 20 Mm (middle row), and 30 Mm (bottom row). The third column shows the variation of the longitudinal (red curve) and transverse (green curve) components of the magnetic field along the line-of-sight for the same three points in the YZ plane in the AR MURaM model. The reference direction for positive Q is parallel to the limb.}
        \label{zeeman-effectQV}
\end{figure*}

\begin{figure*}
    \centering
    \includegraphics[scale=0.28]{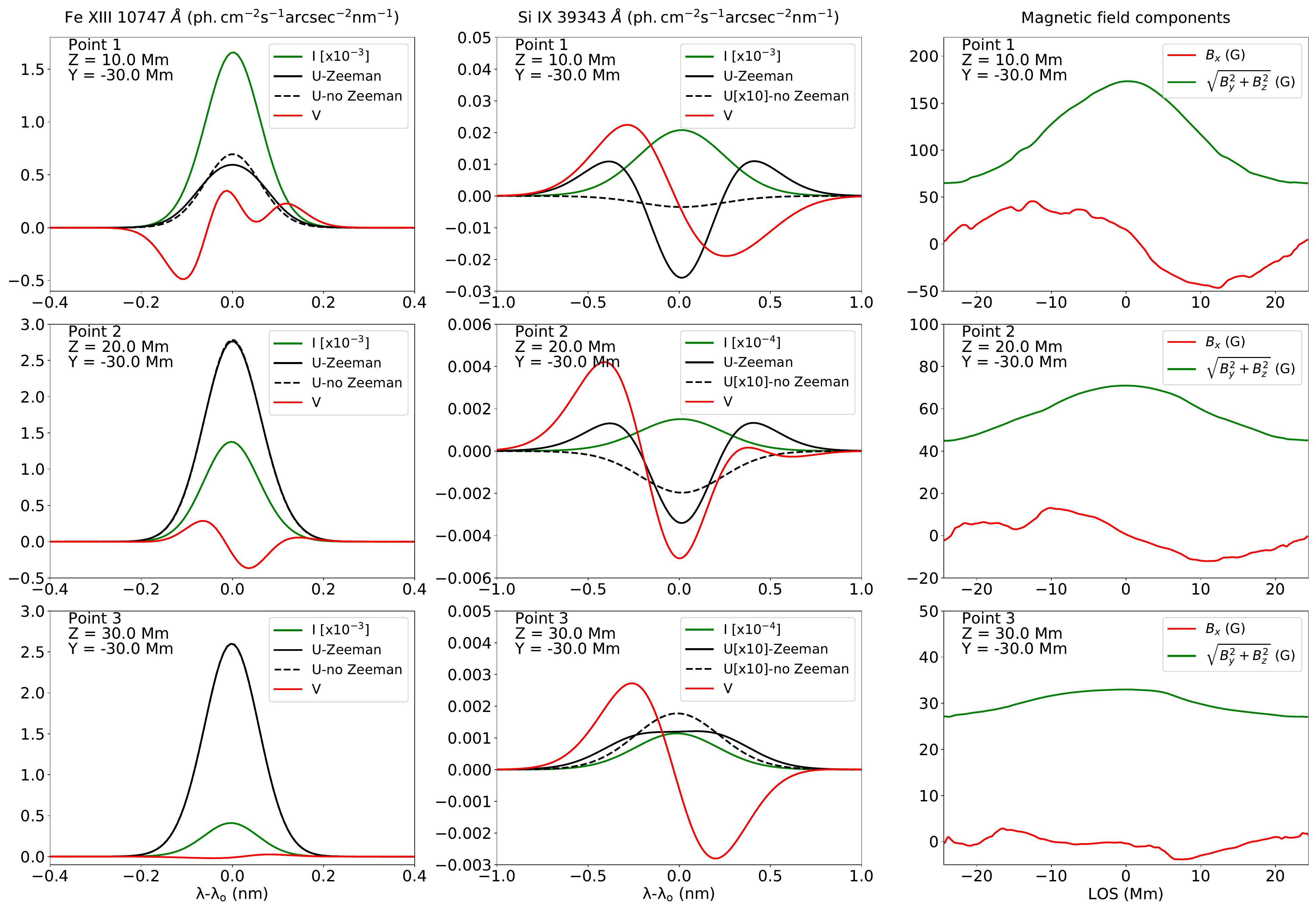}
    \caption{Same as Figure~\ref{zeeman-effectQV} but for Stokes $\rm U$ variation (black curves).}
    \label{zeeman-effectUV}
\end{figure*}

\newpage
\section{Summary and Conclusions} \label{sec:conclusions}
In this paper, we introduced \mbox{P-CORONA}, a forward modeling tool for analyzing the intensity and polarization of spectral lines in 3D solar coronal models. We detailed the structure of the code, the theoretical formulation, and the parallelization methodology adopted. \mbox{P-CORONA} is based on the complete frequency redistribution theory of spectral line polarization, accounting for radiative and collisional transitions in a multi-level atomic system. It includes the effects of scattering due to anisotropic radiation pumping, magnetic fields through the Hanle and Zeeman effects, and non-radial solar wind velocities. The theoretical framework implemented in \mbox{P-CORONA} enables spectral synthesis of both permitted and forbidden lines under various coronal conditions, offering insights into the effects of magnetic and velocity fields on the line polarization. For the magnetic fields expected in the corona, some permitted lines (e.g., H {\sc i} Ly-$\alpha$, O {\sc vi} 
1032 \AA{}, and Ne {\sc viii} \mbox{770 \AA{}}) are generally sensitive to the Hanle effect, whereas forbidden lines, due to the long radiative lifetimes of their levels, typically fall into the saturated Hanle regime and become primarily sensitive to the orientation of the magnetic field. \mbox{P-CORONA} also provides an option for spectral synthesis in the saturated Hanle regime, enabling faster computations when needed.

We demonstrated the capabilities of \mbox{P-CORONA} by applying it to the spectral synthesis of a series of forbidden lines using two sets of models: MURaM models reaching up to coronal heights and large-scale Predictive Science models. The linear and circular polarization signals in the lower heights of the solar corona, as modeled in MURaM, are weak, especially in certain solar conditions and would be challenging to measure them with currently available facilities. However, the results presented here can serve as a reference for future measurements and also help us understand the intensity and polarization signals of different spectral lines under different solar conditions. The spectral synthesis of forbidden lines highlights the complex interplay between physical parameters such as electron temperature, density, and magnetic field. One of the features of \mbox{P-CORONA} is its capability to consistently include the Zeeman effect without relying on the weak field approxima
tion. This allows us to account for the Zeeman effect caused by 
the transverse and longitudinal components of the magnetic field, in relevant cases. We investigated the impact of the Zeeman effect from the transverse magnetic field component in the \mbox{Fe {\sc xiii} 10747 \AA{}} and Si {\sc ix} 39343 \AA{} lines. While signatures of the Zeeman effect are visible at lower heights in the MURaM AR model, the linear polarization signals are too weak to be measurable with current instruments.

In the future, we aim to expand the range of spectral lines analyzed and incorporate additional physical effects, such as symmetry breaking caused by active regions on the photosphere and Thomson scattering. Overall, \mbox{P-CORONA} represents a significant advancement in modeling solar coronal lines, providing a valuable tool for both theoretical and observational studies.

\section{Acknowledgments}
We thank the referee for constructive suggestions, which helped improve the clarity of the manuscript. We also thank Matthias Rempel (HAO) for providing the MURaM models and Tom Schad (NSO) for scientific discussions. 
We acknowledge the funding received from the European Research Council (ERC) under the European Union's Horizon 2020 Research and Innovation Programme (ERC Advanced Grant agreement \mbox{No.~742265}), as well as the support from the Agencia Estatal de Investigación del Ministerio de Ciencia, Innovación
y Universidades (MCIU/AEI) under grant “Polarimetric Inference of Magnetic Fields” and the European
Regional Development Fund (ERDF) with reference
PID2022-136563NB-I00/10.13039/501100011033. T.d.P.A.'s participation in the publication is part of the Project RYC2021-034006-I, funded by MICIN/AEI/10.13039/
501100011033, and the European Union ``NextGenerationEU''/RTRP. N.G.S. is grateful to the Fundación Occident for funding two working visits at the IAC. 
This research was awarded time at the Piz Daint supercomputer by the SOLARNET Trans-national Access Programme, thanks to financial support from the European Union’s Horizon 2020 research and innovation program under grant agreement No. 824135 (SOLARNET).

\clearpage\newpage
\bibliography{AAS62915R2_new.ms}{}
\bibliographystyle{aasjournal}

\end{document}